\def\ls{\lower4pt\hbox{${\buildrel < \over \sim}$}}
\def\gs{\lower4pt\hbox{${\buildrel > \over \sim}$}}

\documentclass[12pt, preprint]{aastex}

\slugcomment{Submitted to {\it The Astrophysical Journal Letters}}

\shorttitle{VHE $\gamma$-Ray Emission in High-z TeV Blazars}
\shortauthors{B\"ottcher, Dermer \& Finke}

\begin{document}

\title{The Hard VHE $\gamma$-Ray Emission in High-Redshift TeV Blazars:
Comptonization of Cosmic Microwave Background Radiation in an Extended 
Jet?}

\author{Markus B\"ottcher\altaffilmark{1}, Charles D. Dermer\altaffilmark{2},
and Justin D. Finke\altaffilmark{2,3}}

\altaffiltext{1}{Astrophysical Institute, Department of Physics and Astronomy, \\
Clippinger 339, Ohio University, Athens, OH 45701; boettchm@ohio.edu}
\altaffiltext{2}{Space Science Division, Code 7653, Naval Research Laboratory, \\
Washington, DC 20375-5352}
\altaffiltext{3}{NRL/NRC Research Associate}

\begin{abstract}
Observations of very-high-energy (VHE, $E > 250$~GeV) $\gamma$-ray 
emission from several blazars at $z > 0.1$ have placed stringent 
constraints on the elusive spectrum and intensity of the intergalactic 
infrared background radiation (IIBR). Correcting their observed VHE
spectrum for $\gamma\gamma$ absorption even by the lowest plausible
level of the IIBR provided evidence for a very hard (photon spectral
index $\Gamma_{\rm ph} < 2$) intrinsic source spectrum out to TeV 
energies.
Such a hard VHE $\gamma$-ray spectrum poses a serious challenge 
to the conventional synchrotron-self-Compton interpretation of the VHE 
emission of TeV blazars and suggests the emergence of a separate
emission component beyond a few hundred GeV. Here we propose that 
such a very hard, slowly variable VHE emission component in TeV blazars 
may be produced via Compton upscattering of Cosmic Microwave Background 
(CMB) photons by shock-accelerated electrons in an extended jet. 
For 
the case of 1ES~1101-232, this component could dominate the bolometric 
luminosity of the extended jet if the magnetic fields are of the order of 
typical intergalactic magnetic fields ($B \sim 10 \, \mu$G) and electrons 
are still being accelerated out to TeV energies ($\gamma \gtrsim 4 \times 
10^6$) on kiloparsec scales along the jet.
\end{abstract}

\keywords{galaxies: active --- BL~Lac Objects: individual (1ES~1101-232) 
--- gamma-rays: theory --- radiation mechanisms: non-thermal}  

\section{Introduction}

More than a dozen blazars have now been detected as sources 
of VHE $\gamma$-ray emission. Almost all of these objects 
are high-frequency-peaked BL~Lac objects (HBLs) at low 
redshifts ($z \lesssim 0.2$). The low redshifts can be 
attributed to the absorption of VHE $\gamma$-rays by 
the IIBR. However, the level of the IIBR in the local 
universe and its evolution with redshift are very difficult 
to measure directly \citep[e.g.,][]{hauser01}. Consequently, 
the intensity and spectral energy distribution of the IIBR and 
the resulting optical depth of the Universe to $\gamma\gamma$ 
absorption and pair production are still a matter of intensive 
debate \citep[see, e.g.,][]{dk05,aha06,sms06,ss06,mr07,aha07}.

The $\gamma\gamma$ opacity of the Universe for GeV -- TeV photons
is generally increasing with increasing photon energy and with
increasing redshift. Therefore, any observed VHE $\gamma$-ray
spectrum with a local photon spectral index $\Gamma_{\rm obs}$ (for 
$\Phi_{\rm obs} (E)$ [ph/(cm$^2$~s)] $\propto E^{- \Gamma_{\rm obs}}$)
corresponds to an intrinsic spectral index $\Gamma_{\rm int}
< 
\Gamma_{\rm obs}$ at the source, i.e., the intrinsic VHE $\gamma$-ray
emission at the source must not only be more luminous, but its spectrum
must also be harder than the observed one. A strict upper limit on
the intensity of the IIBR at the maximum of its spectral energy
distribution (SED) around 1.5~$\mu$m of $\sim 14$~nW~m$^{-2}$~sr$^{-1}$
has been inferred by \cite{aha06} based on the argument that any
plausible mechanism of particle acceleration and radiative cooling
will result in an effective particle spectral lindex $p \ge 2$
(where $n_e (\gamma) \propto \gamma^{-p}$). For any plausible radiation 
mechanism (most notably Compton scattering), this would correspond to a 
limiting intrinsic photon spectral index of $\Gamma_{\rm int} \ge 1.5$. 
According to \cite{aha06} \citep[see also,][]{mr07} the upper limit on
the 1.5~$\mu$m IIBR intensity from the requirement $\Gamma_{\rm int}
\ge 1.5$ for the case of 1ES~1101-232 is now rather close to the lower
limit from direct galaxy counts \citep[e.g.][]{mp00}. Requiring that 
$\Gamma_{\rm int} > 2$ \citep{dermer07} allows a low intensity IIBR 
that is compatible with observations \citep{hauser01}. However, recent 
numerical simulations by \cite{stecker07} indicated that particle spectra 
with $q < 1.5$ can result from diffusive shock acceleration at relativistic 
shocks, which loosens the constraints on the IIBR inferred by \cite{aha06}.
In this context, it should also be mentioned that acceleration at
shear flows \citep[e.g.,][]{so02,rd06} as well as second-order
Fermi acceleration \citep[e.g.][]{vv05} may lead to ultrarelativistic
particle acceleration into particle distributions as hard as $q \sim 1$
or even injection of particle distributions with low-energy cutoffs at 
ultrarelativistic energies \citep{derishev03,sp08}. Subsequent radiative 
cooling will then produce relativistic electron spectra with a canonical 
index of $p = 2$. 

Based on the IIBR spectrum inferred from the constraints described
above, \cite{aha07} reconstructed the intrinsic VHE $\gamma$-ray 
spectrum of the HBL 1ES~1101-232 at $z = 0.186$, which has recently 
been detected as a VHE $\gamma$-ray source by the H.E.S.S. array 
\citep{aha06}. Their reconstruction was based on the limiting value
of the intrinsic spectral index of $\Gamma_{\rm int} \sim 1.5$, 
corresponding to $p \sim 2.0$ (for Compton emission in the Thomson 
limit). Even for the minimum plausible value of the IIBR intensity, 
they inferred that the peak of the high-energy component of 
the SED must be located at $E_p \ge 3$~TeV. While the 
broadband SEDs of HBLs detected at VHE $\gamma$-rays can generally 
be fit with basic synchrotron-self-Compton (SSC) models 
\citep[see, however][for PKS 2155-304]{fdb08}, they are 
very hard-pressed to reproduce such an extremely hard GeV -- TeV 
spectrum \citep[see, e.g.,][]{aha07}. This is because at those 
energies the SSC spectrum exhibits strong intrinsic curvature 
due to Klein-Nishina effects: the number of soft photons that 
can efficiently be scattered (in the Thomson regime) steadily 
decreases with increasing scattered ($\gamma$-ray) photon 
energy. Therefore, even an electron spectrum with $p = 2$ 
will generally produce an SSC spectrum with local spectral 
index at GeV -- TeV energies of $\Gamma_{\rm int} > 1.5$.
One way to overcome this problem could be a very high low-energy 
cutoff of the electron energy distribution ($\gamma_{\rm min} \gtrsim
10^5$), as recently suggested by \cite{kat06}. 
However, maintaining such a high-energy cutoff over the
relevant radiative time scale would require that the emission
region is radiatively inefficient, which is unlikely.

The extremely hard intrinsic GeV -- TeV spectrum of 1ES~1101-232
might suggest that there is an additional, very hard spectral 
component emerging above the SSC emission at those photon 
energies. Here we propose that such a component could be produced
through Compton upscattering of the cosmic microwave background (CMB)
radiation in the extended region of the jet. This mechanism has
previously been considered as a way to explain the
hard X-ray emission in the extended jets of several radio 
quasars and radio galaxies resolved with {\it Chandra} 
\citep[e.g.,][]{sambruna04}, see however \citet{ad04}. The X-ray 
spectra of many of those extended jet components are very hard 
with $\Gamma_x \sim 1.5$, suggesting that even in the case of 
mildly beamed jets of non-blazar radio-loud AGN such 
emission might extend into the MeV -- GeV range. It seems
then worthwhile to investigate whether in the case
of 
blazars, such an emission component could even extend into
the GeV -- TeV regime and produce a quasi-steady plateau of
very hard VHE emission. 

Throughout this paper, we refer to $\alpha$ as the energy 
spectral index, $F_{\nu}$~[Jy]~$\propto \nu^{-\alpha}$ and
the photon spectral index $\Gamma_{\rm ph} = \alpha + 1$. A 
cosmology with $\Omega_m = 0.3$, $\Omega_{\Lambda} = 0.7$, 
and $H_0 = 70$~km~s$^{-1}$~Mpc$^{-1}$ is used. In this cosmology,
and using the redshift of $z = 0.186$, the luminosity distance 
of 1ES~1101-232 is $d_L = 905$~Mpc. 

\section{\label{estimates}Compton upscattering of the CMB to
VHE $\gamma$-rays}

The CMB at the source at redshift $z$ is a thermal blackbody 
spectrum with a temperature $T_{\rm CMB} = (1 + z) \times 
2.72$~K. Its local energy density is
$u_{\rm CMB} = 
(1 + z)^4 \times 4.02 \times 10^{-13}$~ergs~cm$^{-3}$.
For the purpose of an analytical estimate, the CMB can be
approximated as a monochromatic radiation field at a normalized mean
photon energy $\epsilon_{\rm CMB} \equiv (h \nu_{\rm CMB} / m_e c^2) 
\approx (1 + z) \times 1.2 \times 10^{-9}$. If the emission 
region is moving with respect to the local AGN frame with a 
bulk Lorentz factor $\Gamma$, the CMB radiation field is 
boosted into the comoving frame by a factor of $\Gamma$ in photon
energy and a factor of $\Gamma^2$ in energy density. In the following,
we scale $\Gamma = 10 \, \Gamma_1$, and, accordingly, the Doppler
boosting factor $D \equiv \left( \Gamma [1 - \beta_{\Gamma}
\cos\theta_{\rm obs}] \right)^{-1} \equiv 10 \, D_1$, where
$\theta_{\rm obs}$ is the observing angle. Primed
quantities 
refer to the co-moving frame of the emission region.

Efficient Compton upscattering of the CMB into the $\gamma$-ray
regime with the observed very hard intrinsic spectrum of 1ES~1101-232,
requires that the scattering occurs in the Thomson regime which 
is possible for electrons with Lorentz factors of $\gamma' 
\le (4 \Gamma \epsilon'_{\rm CMB})^{-1} = 2.1 \cdot 10^7 \,
\Gamma_1^{-1} \, (1 + z)^{-1}$. Such electrons will Compton
scatter CMB photons up to energies of $\epsilon_{\rm s, max}^{\rm obs}
\approx 5.3 \times 10^7 \, D_1 \, \Gamma_1^{-1} \, (1 + z)^{-2}$
or $E_{\rm s, max}^{\rm obs} \approx 27 \, D_1 \, \Gamma_1^{-1} \, 
(1 + z)^{-2}$~TeV in the observer's frame. Thus, if electrons can 
be accelerated to TeV energies, they will be able to upscatter CMB 
photons into the TeV regime in the Thomson limit. Specifically,
to scatter a CMB photon up to 1~TeV in the observer's frame, a
comoving electron Lorentz factor of $\gamma'_{\rm TeV} = 4.0 
\times 10^6 \, (D_1 \Gamma_1)^{-1/2}$ is required. The radiative 
cooling time scale due to Compton cooling in the co-moving frame is 

\begin{equation}
\tau'_{\rm CMB} (\gamma_{\rm TeV}) = 6.3 \times 10^3 \, D_1^{1/2} \, 
\Gamma_1^{-3/2} \, (1 + z)^{-1} \; {\rm yr}.
\label{taucmb}
\end{equation}

Based on this cooling time scale it will in principle be 
possible to accelerate electrons out to $\gamma'_{\rm TeV}$ 
for magnetic field values exceeding $B' \gtrsim 1.2 \times 10^{-12} \, 
D_1^{-1} \, \Gamma_1 \, (1 + z)^4$~G, which does not
impose a serious constraint. CMB Compton cooling will dominate over 
synchrotron cooling for a co-moving magnetic field value of 

\begin{equation}
B' \lesssim 3.2 \times 10^{-5} \, \Gamma_1 \, (1 + z)^2 \; {\rm G}. 
\label{Bmax}
\end{equation}

The resulting synchrotron emission produced by electrons with 
energy $\gamma'_{\rm TeV}$ will peak at $\nu_{\rm sy}^{\rm obs}
\lesssim 2.19 \times 10^{16} \, (1 + z)$~Hz, i.e., typically
in the UV or soft X-ray bands. The dominance of CMB Compton 
cooling over synchrotron cooling is required in order for 
the CMB Compton emission to be at least at a comparable 
level to the synchrotron emission from the same electrons. 
The required low magnetic fields are much lower than the 
magnetic field strengths inferred for the presumably
near-central (sub-pc) regions of the jet from where the bulk 
of the observed optical -- X-ray -- GeV $\gamma$-ray emission 
from blazars is generally believed to originate. This points
towards an origin in the outer (pc -- kpc scale) regions of
the jet, where the magnetic fields might approach characteristic
interstellar magnetic field values of a few tens of $\mu$G. 
The resulting magnetic jet luminosity will be $L_{\rm mag}
\sim 1.5 \times 10^{39} \, R_{18}^2 \, \Gamma_1^4$~ergs~s$^{-1}$,
where $R_{18}$ is the (poorly constrained) cross-sectional 
radius of the jet in units of $10^{18}$~cm, and thus several 
orders of magnitudes less than the total luminosity required 
to power the emission. This indicates that the jets must be 
particle dominated at the site of the VHE $\gamma$-ray production 
through CMB upscattering, even for a pc-scale transverse extent 
of the jet.

A key prediction from the inferred long cooling time scale 
given above is that the CMB Compton emission, as well as the
synchrotron emission associated with the same emission region,
will be slowly variable. This is another reason why the 
magnetic-field energy density in the VHE emitting region 
should be lower than the CMB energy density: otherwise
this would contradict the observations of rapid variability 
throughout the synchrotron component (in particular, at optical
and X-ray frequencies) of most blazars, including 1ES~1101-232
\citep[e.g.,][]{remillard89,romero99,wolter00}. We therefore
argue that the rapidly variable synchrotron emission originates
on substantially smaller (sub-pc) scales along the jet than the
VHE emission.

\begin{figure}
\plotone{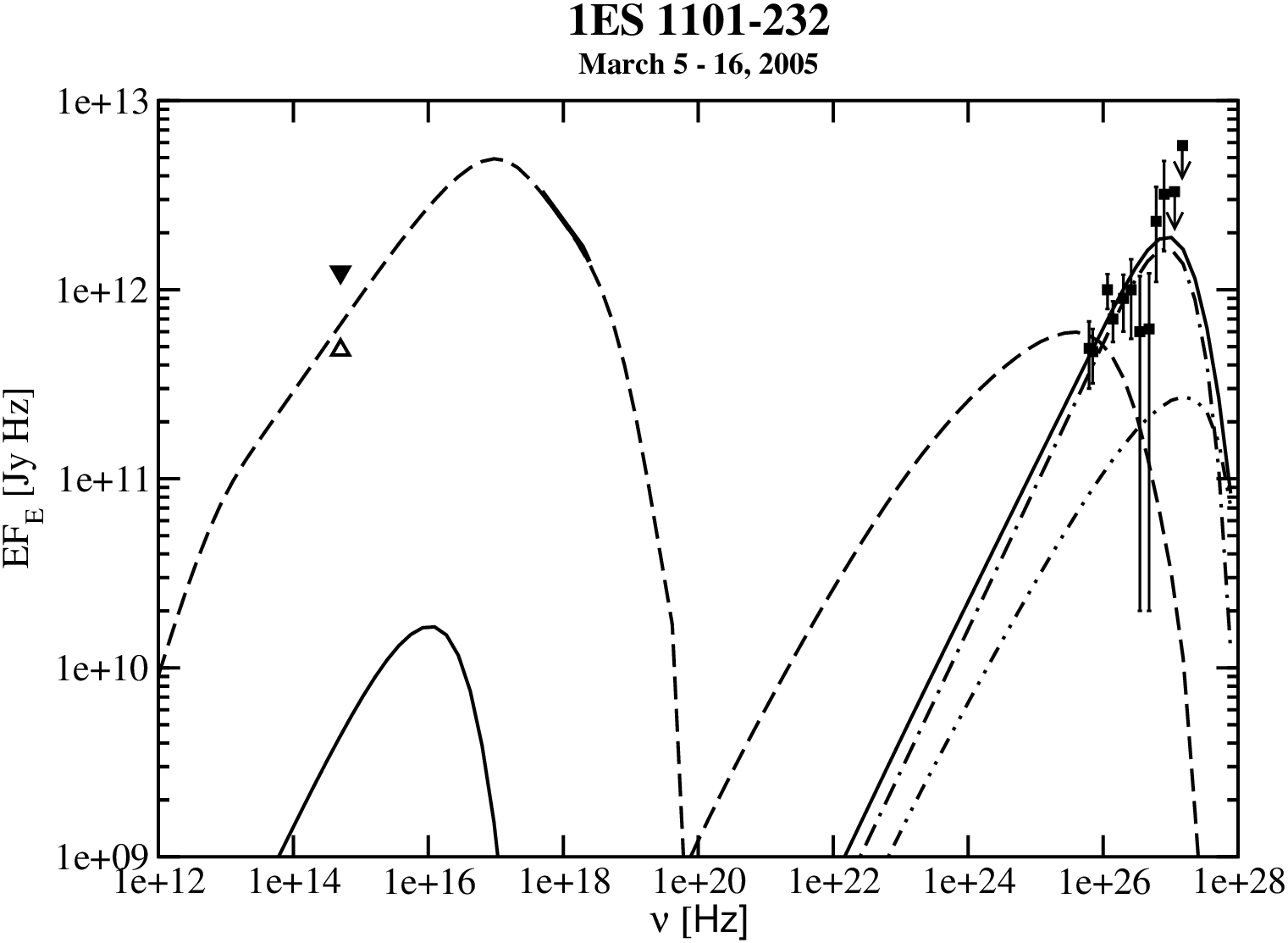}
\caption{Simultaneous optical -- X-ray -- VHE $\gamma$-ray SED 
of 1ES~1101-232 on March 5 -- 16, 2005 \citep[from][]{aha07}. 
The VHE measurements have been corrected for absorption by the
IIBR as described in \cite{aha06}. The optical data points are
an upper and a lower R-band limit, respectively, derived from 
simultaneous ROTSE~3c observations in the 400 -- 900 nm bandpass 
\citep{aha07}. The curves indicate a model fit using an extended 
version of the code of \cite{bc02}. The H.E.S.S. data have been 
fitted with $\Gamma = 15$, $B' = 10 \, \mu$G, ${\gamma'}_1 = 100$, 
${\gamma'}_2 = 6.0 \times 10^6$, $q = 1.5$, and $L_{\rm jet} 
= 1.5 \cdot 10^{41}$~ergs~s$^{-1}$. 
The individual curves represent: dot-dashed = CMB Compton; 
double-dot-dashed = SSC; solid = total emission from the 
VHE emitting region. The dashed curve, fitting the synchrotron
component, has been computed with parameters more appropriate
to the inner-jet region, with $\Gamma = 25$, ${\gamma'}_1 = 1.3 
\times 10^5$, ${\gamma'}_2 = 7.5 \times 10^5$, $q = 3.2$, 
$L_{\rm jet} = 2.0 \times 10^{40}$~ergs~s$^{-1}$, $R'_B = 
6 \cdot 10^{16}$~cm, and a magnetic field in equipartition
with the relativistic electron plasma, at $B' = 0.06$~G. 
The X-ray/$\gamma$-ray portion of the dashed curve represents
the inner-jet SSC emission.}
\label{1101_march05}
\end{figure}

\section{\label{results}Results for 1ES~1101-232}

\begin{figure}
\plotone{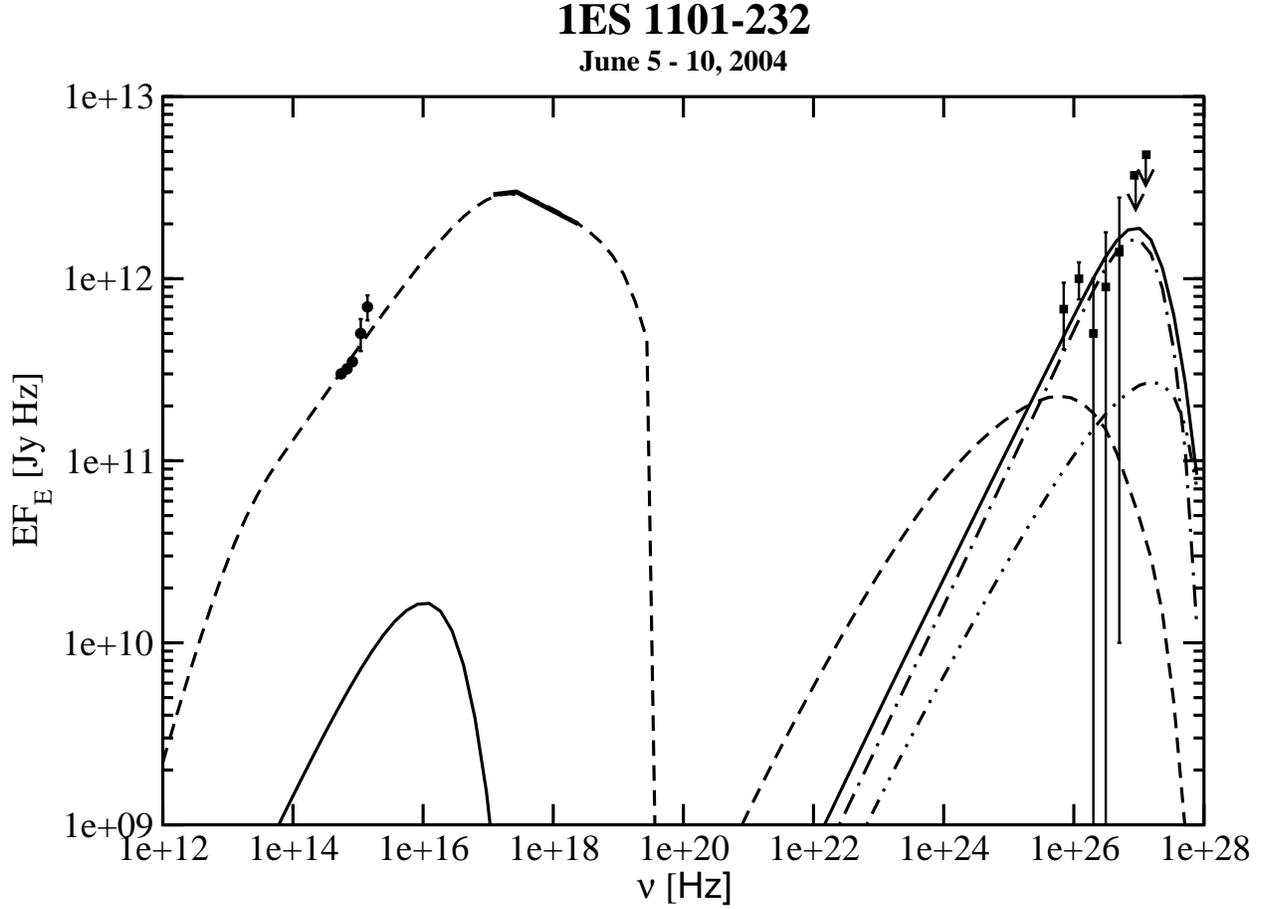}
\caption{Same as Fig. \ref{1101_march05}, but for the SED of
June 5 -- 10, 2004. Here, optical and X-ray data were obtained
from simultaneous {\it XMM-Newton} observations \citep{aha07}.
Inner-jet parameters used for the synchrotron fit were: $\Gamma = 25$, 
${\gamma'}_1 = 1.8 \times 10^5$, ${\gamma'}_2 = 1.5 \times 10^6$, 
$q = 2.5$, $L_{\rm jet} = 1.45 \times 10^{40}$~ergs~s$^{-1}$, 
$R'_B = 6 \cdot 10^{16}$~cm, and $B' = 0.05$~G.}
\label{1101_june04}
\end{figure}

Based on the similarity of the optical and X-ray characteristics 
of 1ES~1101-232 \citep{remillard89,falomo94} with
previously 
known TeV blazars, \cite{wolter00} and \cite{cg02}
predicted 
that 1ES~1101-232 might be detectable by the now
operating 
generation of VHE $\gamma$-ray detectors. Prompted
by 
these predictions, \citet{aha06} performed three sets of
coordinated optical -- X-ray -- VHE $\gamma$-ray observations 
in 2004 and 2005, which resulted in its detection at VHE
$\gamma$-rays. This made 1ES~1101-232 the highest-redshift
BL~Lac object detected in VHE $\gamma$-rays at the time (now
surpassed by 1ES~1011+496 at $z = 0.212$, \cite{albert07}) 
and the second-farthest
VHE $\gamma$-ray source known, 
only topped by the quasar-type blazar 3C~279 at 
$z = 0.538$, for which the MAGIC collaboration
recently claimed a VHE detection \citep{tes07}. The H.E.S.S.
observations of 1ES~1101-232 provided no evidence for variability
at VHE $\gamma$-rays. Figs. \ref{1101_march05} and \ref{1101_june04} 
show the simultaneous optical -- X-ray -- VHE $\gamma$-ray spectra
during two of the three H.E.S.S. observing windows, where the
VHE $\gamma$-ray points have been corrected for $\gamma\gamma$
absorption by the IIBR as described in \cite{aha07}.

Unfortunately, all of the known observational facts on
1ES~1101-232 do not allow very meaningful constraints on 
the parameters of the kpc-scale jets of 1ES~1101-232 beyond 
a constraint on the overall jet power: The power injected 
into the jet on kpc scales obviously needs to exceed the
observed radio power of the extended radio structure at
1.5~GHz, $P_{\rm 1.5}^{\rm ext} \approx 3.8 \times 10^{40}$~erg/s
\citep{lm93}. Following arguments presented for the X-ray emission 
from
extended jets of other AGN, indicating that the jets remain
highly relativistic out to kpc scales \citep[e.g.][]{tav00,ad04,schwartz06}, 
we assume a characteristic Lorentz factor of $\Gamma = D = 15$.
Based on the discussion in the previous section, an 
extension of the CMB Compton emission into the TeV
regime requires electrons with energies up to $\gamma'_2
\gtrsim 2.7 \cdot 10^6$. In order for the (non-variable)
synchrotron emission from the TeV emission region not to
dominate the observed X-ray emission, a magnetic field of
$B' \lesssim 675 \, \mu$G is required. 

In order to evaluate the resulting broadband spectrum from
a relativistic jet component, we have incorporated the process
of CMB Compton scattering into the jet radiation transfer code
described in detail in \cite{bc02}. We used the head-on approximation
for Compton scattering in the Thomson regime to treat the Compton
scattering process and describe the CMB as a thermal blackbody
at $T_{\rm CMB} = 2.72 \, (1 + z)$~K. 

A reasonable representation of the VHE $\gamma$-ray spectrum
as shown in Fig.\ \ref{1101_march05} can be achieved using an 
electron injection index of $q = 1.5$, a luminosity injected 
in relativistic electrons into the kpc-scale jet of $L_{\rm jet} 
= 1.5 \times 10^{41}$~ergs~s$^{-1}$, and a co-moving magnetic 
field of $B' = 10 \, \mu$G. Obviously, these parameters are not 
tightly constrained, given that only the rather small spectral 
range of the VHE spectrum is fitted. The fits are insensitive
to a low-energy cutoff ${\gamma'}_1$ of the electron spectrum, at 
least as long as ${\gamma'}_1 \lesssim 10^3$. The same choice of
parameters provides an equally acceptable fit to the SED of
June 5 -- 10, 2004, as shown in Fig. \ref{1101_june04}.

As expected (and required), the predicted synchrotron emission 
from the extended jet region is about two orders of magnitude 
below the measured SED for both observation periods. In order
to provide a fit to the synchrotron component of 1ES~1101-232,
we chose parameters more appropriate for the inner (sub-pc) 
region of the jet, as listed in the figure captions. Note that
the jet powers quoted there refer only to the power carried
by the ultrarelativistic particle content. In the fits to the
synchrotron component, we have chosen the magnetic field in
equipartition with the ultrarelativistic electron population
in the emission region in order to reduce the number of free
parameters. 

Although the parameters for the fits to both the synchrotron
and the VHE components are not very well constrained, the general 
picture that emerges is that the CMB-Compton interpretation
for the VHE emission requires a substantially larger kinetic 
luminosity of ultrarelativistic electron injection into 
the emission region on large scales than inferred for the 
synchrotron component at sub-pc scales. This suggests that 
a substantial fraction of the total jet luminosity is carried
out to large distances by the non-leptonic content of the jet.
If our assumption of a magnetic field in equipartition with
relativistic electrons on sub-pc scales is realistic, this
would suggest that most of the kinetic power of the jet might
be carried by hadrons and dissipated on pc -- kpc scales, 
leading to the acceleration of ultrarelativistic electrons in 
the extended jet. This is in perfect agreement with results 
of several other authors, suggesting that the pc-scale jets
of blazars are dominated in number by electron-positron pairs,
responsible for the high-energy emission produced at sup-pc
scales, while the bulk of the kinetic power of the jets 
is carried by a relativistic hadron population 
\citep[e.g.][]{sm00,gc01,kt04,sikora05}. 

\section{\label{discussion}Summary and Discussion}

We have presented a possible explanation for the hard 
VHE spectra of high-redshift $\gamma$-ray blazars, when 
corrected for the expected amount of $\gamma\gamma$ 
absorption by the IIBR. We suggest that ultrarelativistic 
electrons may be accelerated on large (kpc) scales, 
where magnetic fields are so low that synchrotron 
cooling becomes negligible compared to Compton 
cooling on the CMB. Consequently, VHE emission 
from Compton upscattering of the CMB by ultrarelativistic 
electrons may produce a separate, slowly variable VHE 
component beyond the rapidly variable broadband 
continuum from radio through GeV $\gamma$-ray 
energies that might be dominated by emission 
from the sub-pc scale, inner regions of the jet. 

As of the time of writing of this manuscript, the VHE
emission of most distant TeV blazars, e.g., 1ES~1101-232 
\citep{aha07}, 1ES~0229+200 \citep{aha07c}, 1ES~1011+496 
\citep{albert07}, has not shown evidence for variability
on any time scale, while the nearby TeV blazars show 
substantial variability on time scales from years down
to a few minutes \citep[e.g.][]{aharonian07b}. This could
indicate that CMB Compton scattering constitutes an underlying
non-varying VHE component in TeV blazars, which would be more
likely to be detected in high-redshift sources since the energy 
density of the CMB increases with redshift as $(1 + z)^4$. 
However, one has to caution that small-amplitude variability 
would generally be much harder to detect for the much fainter 
high-redshift TeV sources.

We have applied this idea to two simultaneous SEDs of 1ES~1101-232.
Although the parameters of our fits are rather poorly constrained,
they seem to indicate that only a small portion of the kinetic
energy of the jet is dissipated on sub-pc scales, while the bulk
of this energy is transported outward to kpc scales, in agreement
with earlier suggestions that the kinetic power of blazar jets
might generally be dominated by their hadronic content and is only
dissipated on large scales. 

Alternative explanations for the very hard VHE emissions include
Compton upscattering of infrared emission from dust in the
vicinity of the AGN \citep[e.g.,][]{blaz00}, or signatures of 
pion decay in hadronic blazar models \citep[e.g.,][]{muecke03}. 
However, it should be noted that Compton scattering of infrared 
radiation with a characteristic wavelength of $\lambda = 
\lambda_1 \, \mu$m would also be affected by Klein-Nishina
effects at energies of $E \gtrsim 0.2 \, \lambda_1$~TeV, so
a photon index of $\Gamma_{\rm int} \sim 1.5$ out to multi-TeV
energies might be difficult to explain with this mechanism.
It is also possible to produce VHE spectra of arbitrary hardness
if they are affected by $\gamma\gamma$ absorption intrinsic to
the source through interaction with narrow-band emission from
the vicinity of the AGN \citep{aha08}. However, it is not
obvious why such a feature should be observed preferentially
in high-redshift TeV blazars, and why it would lead to less 
variable TeV emission than in the low-redshift TeV blazars, 
if the lack of variability is confirmed by future observations. 

Our CMB-Compton interpretation for the VHE emission of 1ES~1101-232
is obviously not applicable to high-redshift VHE sources that 
seem to show substantial variability on time scales of days or 
even shorter, e.g., 3C279 \citep{tes07} since it requires a
pc-scale extent of the emission region. The interpretation
of the unexpectedly hard VHE spectrum through such a scenario
would predict that any emission at lower (GeV) energies might 
be produced in the inner (sub-pc) region of the jet and be 
much more rapidly variable than the VHE emission. Whether 
this is the correct interpretation for the hard spectrum of 
1ES~1101-232 or not, a Compton-scattered CMB component from 
blazar jet emission will accompany synchrotron emission and 
could potentially be observed with $\gamma$-ray telescopes.

\acknowledgments
We thank the referee for a very helpful and constructive
report which helped to improve our manuscript substantially.
The work of M.\ B\"ottcher as an ASEE Summer Research Faculty 
Fellowship and J.\ Finke as an NRC/NRL postdoctoral research 
associate at the Naval Research Laboratory was supported 
by NASA {\it GLAST} Science Investigation DPR-S-1563-Y. 
The work of C.\ D.\ is supported by the Office of 
Naval Research.

\end{document}